%% file: Final19.tex
\documentclass{article}
\usepackage{makeidx, color}
\usepackage{setspace}
\usepackage{enumerate}
\usepackage{subfigure}
\usepackage{delarray}
\usepackage{latexsym}
\usepackage{amsfonts}
\usepackage{array}
\usepackage{amsmath}
\usepackage{amssymb}
\usepackage{amsbsy}
\usepackage{amsgen}
\usepackage{mathrsfs}
\usepackage{lmodern}
\usepackage[dvips]{epsfig}

\title{Mixing time of $A + B -> 0$ in 1D.}
\author{C P Haynes \\
School of Mathematics and Physics, \\
The University of Queensland,\\ St Lucia, Queensland, 4072, Australia.}
\date{}

\begin{document}

\maketitle

\begin{abstract}
A mixing time density of $A + B \to 0$ on a finite one dimensional domain is defined for general initial and boundary conditions in which $A$ and $B$ diffuse at the same rate. The density is a measure of the number of $A$ and $B$ particles that mix through the center of the reaction zone. It also corresponds to the reaction density for the special case in which $A$ and $B$ annihilate upon contact. An exact expression is found for the generating function of the mixing time. The analysis is extended to multiple reaction fronts and finitely ramified fractals. The full method involves using the kernel of the Laplace transform integral operator to map and analyze a moving homogeneous Dirichlet interior point condition. 
\end{abstract}

\section{Introduction}
Comprehension of diffusion-reaction schemes in disordered media is often marred by the symbiotic nature between the diffusive aspect and the reactive aspect of the scheme. If the underlying diffusion is not understood then it is difficult to define the reactivity, whereas the reaction component naturally impedes the diffusivity of the reactants. Consequently, a major goal in experimental and theoretical studies  is to elucidate the diffusive or reactive aspects of a scheme which requires understanding, for example; the effect of crowded intracellular conditions on the reduced mobility of a reactant~\cite{Klann2011,Schoen2011}, the partial differential equation description of a system~\cite{Flegg2013}, or the diffusion limit in inhomogeneous environments~\cite{Lim2014}. 

For the irreversible bimolecular reaction between two different initially separated diffusing species $A + B \to 0$, a significant study of the diffusive and reactive aspects has occurred over the past 30 years~\cite{Richardson1997, Barkema1996, Araujo1993, Kisilevich2008, Shipilevsky2008}. Even so, given that $A+B\to 0$ is one of the most elementary schemes that finds application in the physical~\cite{Toussaint1983, Kwon2006}, biological~\cite{Frachebourg1996} and chemical~\cite{Kopelman1988} sciences, the derivation of new results concerning either the diffusive or reactive aspects is important. The standard theoretical approach in a one-dimensional system involves solving the set of equations~\cite{Avraham2000}  
\begin{eqnarray}\label{eq:siam55}
D_A \partial_{xx} \,C_A(x,t) - \partial_t\, C_A(x,t) & = 
& R(x,t), \nonumber \\
D_B \partial_{xx} \,C_B(x,t) - \partial_t\, C_B(x,t) & =
& R(x,t), \nonumber
\end{eqnarray}
 where $C_A$ and $C_B$ are the concentrations of each species remaining within the system at time $t$, $D_A$ and $D_B$ being the diffusion coefficients of the reactants, $R$ the macroscopic reaction rate and $\partial_x C_A(X,t)$ is the partial derivative of $C_A$ with respect to $x$ evaluated at $x = X$. As time evolves, the $A$ and $B$ species will mix forming a reaction zone whose evolution depends on $R$.

Studies of $A + B\to 0$ can employ approximative forms of $R$ in order to solve the underlying model~\cite{Ben1992}; typically, a mean field density of the form $R = k\,C_AC_B$ ($k$ constant) is introduced thereby making the system nonlinear.  The validation of such an approximation has been confirmed through experimental studies~\cite{Koo1991, Monson2004}, although deviations~\cite{Cornell1993} from the mean-field behavior are known to occur and the exact form for $R$ remains unresolved. Alternatively, by focusing on the diffusive aspect, it is possible to study the encounter rate of $A$ and $B$, which becomes useful in determining an upper bound on the rates of reaction. 

At the locations where $A$ and $B$ encounter one another they begin to mix. The way in which they mix is affected by the initial placement of the species~\cite{Lindenberg1998}. If it happens that both species exhaust themselves in the mixing process, then the mixing is  homogeneous (or efficient), otherwise it is inhomogeneous. For example, on an infinite one dimensional domain, if the total number of $A$ species in a particular region is greater than (and surrounds) the number of $B$ species then a rapid disappearance of the $B$ species will eventuate from in-homogeneous mixing thus resulting in segregated island-type phenomenon~\cite{Shipilevsky2013}. For finite domains, the boundary of the domain will accentuate this mixing behavior. Providing a general condition for when this behavior occurs is essential to the reactive aspect of the $A+B\to 0$ scheme and all schemes having similar theoretical form ( i.e. $A + B \to C$). 

It is possible to infer results pertaining to the mixing of the species, irrespective of $R$, when $D_A = D_B$. This is done by studying a fluctuation density of the system , $C = C_A - C_B$. The point at which $C = 0$, $x = M(t)$ (which is unique, say) represents the point where the $A$ and $B$ species mix. Studying the flux at $x = M(t)$ will obtain the reaction density in the special case that $A$ and $B$ annihilate upon contact, however, there appears to lack a comprehensive theoretical analysis of the flux at $M(t)$, its implications in classifying the mixing behavior of the species and its implications for the reactive part of the scheme.

The flux at the mixing point $x = M(t)$, $F(t)$,  in some sense solves the diffusional aspect of the diffusion-reaction process given that it represents the number of new $A$ and $B$ species that are mixed and are able to react. If the mixing point is defined for all time, then $A$ and $B$ are mixing homogeneously. A study involving the author~\cite{Haynes2012} found criterion for when homogeneous mixing occurred on a finite one dimensional domain with no flux boundaries, although the extension to arbitrary boundary conditions and one dimensional networks (i.e. finitely ramified fractals) has not been done. 

This study finds the generating function of $F$ for general initial and boundary conditions on one dimensional domains, for the case in which $A$ and $B$ are initially separated, diffuse at the same rate and for which the mixing point is defined for all time. To the best of the author's knowledge no such in-depth study has been performed. 
The paper is structured as follows: we define the mathematical problem for a finite one dimensional domain in \S~\ref{sec:check}. The results for the case in which homogeneous mixing occurs are provided within \S~\ref{sec:statphys4}, with the analysis used to obtain these results being presented in \S~\ref{sec:statphys}. Conditions determining when the mixing point $M(t)$ is unique are provided in \S~\ref{sec:statphys2}. In \S~\ref{sec:siam1}, multiple mixing points are considered  and conditions are provided on whether homogeneous  mixing occurs. The extension to one-dimensional networks is then presented in \S~\ref{sec:siam2}. We conclude the study in \S~\ref{sec:statphysend}. 

\section{Finite one dimensional domain}~\label{sec:siam5}

\subsection{Focus}~\label{sec:check}
The analysis involves working with $C = C_A - C_B$, where $C(x,t):  \Omega  \to \mathbb{R} $, $\Omega = [0, \ell] \times [0,\infty)$, is the solution of 
\begin{eqnarray}\label{eq:statphys8}
D \partial_{xx}\, C = \partial_t C,  \quad C(x,0) = \rho_A(x) - v\,\rho_B(x), \nonumber \\
\partial_x \,C(0,t) = J_1(t), \quad \partial_x \,C(\ell , t) = J_2(t), \quad \| J_i \| = \int_0^{\infty} |J_i|\, dt.
\end{eqnarray}
Here, $\rho_A$ and $\rho_B$ are the initial continuous density functions of the $A$ and $B$ species whose support lies in $[I_A^-, I_A^+]$ and $[I_B^-, I_B^+]$ (respectively), where $I_A^- >0, I_B^+ < \ell$, $I_B^- > I_A^+$ and $v \in [0,1]$ is a parameter that sets the initial ratio of $A$ and $B$. $J_1(t)$ and $J_2(t)$ are taken to be within the space of continuous functions. For this type of $C(x,0)$, a moving homogeneous Dirichlet interior point condition is defined where the positive and negative concentrations meet. This occurs at some point $x = M(t)$, such that $C(M(t),t) = 0$, where the movement of $M(t)$ is fully determined by the flux at $x =M(t)$, $F(t)$, since 
\begin{equation}\label{eq:statphys3}
 F(t) = F(t)^- = \lim_{x \to M(t)^-} - D\partial_x \,C \equiv  \lim_{x \to M(t)^+} \left| D\partial_x \,C\right|  = F(t)^+ = F(t).
\end{equation}
Here $M(t)$, which we also refer to as the mixing point, will always occur regardless of the form of $J_i$, however it may not be defined for all time. An illustration of the problem is given in Fig.~\ref{fig:statphys} a) and c). 

\subsection{Results}\label{sec:statphys4}
The goal is to find the generating function of $F(t)$, $f(s)$, which we define through the Laplace transform $f(s) = \mathscr{L}[F(t)] = \int_0^{\infty} \, F(t)\,e^{-st} \, dt,\,(s>0)$. Throughout, we also make use of the Laplace-Stieltjes transform, $\mathscr{L}_S$, where
$
\mathscr{L}_S\left[F \right] = \int_0^{\infty} e^{-st}\,d\left[F(t)\right].
$
The procedure involves first solving the Laplace transformed system of~(\ref{eq:statphys8}), 
\begin{equation}\label{eq:statphys2}
 D\Delta c - s\,c = -\rho_A  + v\rho_B,\quad \partial_x\,c(0,s) = j_1(s), \quad \partial_x\,c(\ell,s) = j_2(s),
\end{equation}
where $c(x,s) = \mathscr{L}[C(x,t)]$, $j_1(s) = \mathscr{L}[J_1(t)]$ and $j_2(s) = \mathscr{L}[J_2(t)]$, to obtain
\begin{equation}\label{eq:statphys4}
c = \frac{\textrm{ch}(\ell-x)}{\textrm{sh}(\ell)\sqrt{sD}}\,\left(j_1 + Z_1[x ,C(x,0)] \right)  +  \frac{\textrm{ch}(x)}{\textrm{sh}(\ell)\sqrt{sD}}\,\left(j_2 + Z_2[x, C(x,0)] \right). \\
\end{equation}
Here $\textrm{sh}(x) = \textrm{sinh}(x\sqrt{\frac{s}{D}})$, $\textrm{ch}(x) = \textrm{cosh}(x\sqrt{\frac{s}{D}})$, $Z_1\left[x, C(x,0) \right] = \int_0^{x} \textrm{ch}(x')\,C(x',0)\,dx'$,  $Z_2\left[x, C(x,0) \right] = \int_x^{\ell} \textrm{ch}(\ell-x')\,C(x',0)\,dx'$, with the above representation being in the Green's function form~\cite[Pp253]{Hunter2001}. In \S~{\ref{sec:statphys}, it is shown that there exists a point $x = y(s)$ such that 
\begin{displaymath}
 f(s) = D \partial_x\,c(y(s),s) = \mathscr{L}\left[D \partial_x C(M(t),t)\right]  = \mathscr{L}[F(t)],
\end{displaymath}
where
\begin{equation}\label{eq:statphys9}
y(s) = \frac{1}{\sqrt{s}}\,\textrm{arctanh}\left(\textrm{sh}(\ell)^{-1}\left(\textrm{ch}(\ell) -  \frac{vZ_2\left[I^-_B, \rho_B \right] - j_2 }{j_1 + Z_1\left[I^+_A, \rho_A\right]}\right)\right),
\end{equation}
and
\begin{equation}\label{eq:statphys10}
f = \sqrt{\left(Z_1\left[I^+_A, \rho_A\right]+j_1\right)^2   - \left(\frac{\left(Z_1\left[I^+_A, \rho_A\right]+j_1\right) \textrm{ch}(\ell) - vZ_2\left[I^-_B, \rho_B\right] + j_2}{\textrm{sh}(\ell)}\right)^2}.  \\ 
\end{equation}
These results are valid provided $M(t)$ is defined for all time, and is the unique point in which $C(M(t),t) = 0$. A uniqueness criteria for $M(t)$ is given in \S~{\ref{sec:statphys2}}. When $M(t)$ is not defined for all time, an alternative method is required. Note that $j_i$ can be set to consider Dirichlet boundary conditions, given the inter-relation of $j_1$, $j_2$, $c(0,s)$ and $c(\ell,s)$ found by substituting $x = 0$ and $x = \ell$ in Eq.~(\ref{eq:statphys4});
$$
c(\ell,s) =  \frac{\textrm{ch}(\ell)\,j_2  + j_1 + Z_1\left[\ell, C(x,0)\right]}{\sqrt{sD}\,\textrm{sh}(\ell)} ,  \quad c(0,s) =  \frac{\textrm{ch}(\ell)\,j_1  + j_2 + Z_2\left[0, C(x,0) \right]}{\sqrt{sD}\,\textrm{sh}(\ell)}.
$$
The results have been numerically validated for an example (see Fig.~\ref{fig:statphys} b)).

\subsection{Analysis}\label{sec:statphys}
The purpose of this section is to prove that $\partial_x C(M(t),t) = \mathscr{L}^{-1}[\partial_x c(y(s),s)]$.

$C(x,t)$ is a scalar function of time and space, but can alternatively be defined by letting $x$ assume a  function of time; $C(X(t),t)$, where $X$ belongs to a set $\mathbb{B}_t$ of moving points defined for all $t \geq 0$. If some $\kappa_t \subset \mathbb{B}_t$ is to span $\Omega$ such that every $X(t) \in \kappa_t$ is distinct for each $t$, we require $X(t)$ to be a translation of a designated origin point $X_0(t)$ within $\kappa_t$:
\begin{equation}\label{eq:origin}
X(t) = [(X_0(t) + \xi) \,\textrm{mod}\, \ell] \,\textrm{mod} \,0, \quad \xi \in [0,\ell].
\end{equation}
This implies that there are moving points which possess discontinuities in time as they cross the boundary (i.e. at some time $t_c$, $\lim_{t \to t^-_c} X(t) = \ell$, $\lim_{t \to t^+_c} X(t) = 0$) and that 
$$ 
\int_0^\ell C(x,t)\,dx = \int_0^\ell C\left(\left[\left(X_0(t) + \xi\right) \,\textrm{mod}\, \ell\right]\, \textrm{mod}\, 0,t\right) \,d\xi \quad \forall t \geq 0.
$$
Analogously, $c(x,s)$ can be considered in terms of $c(Y(s),s)$. Here $Y \in \mathbb{B}_s$, which is a set of moving points defined for all $s \geq 0$ such that if some $\kappa_s \subset  \mathbb{B}_s$ spans $\Omega$, with every $Y(s)\in \kappa_s$ being a translation of a designated origin point $Y_0(s)$, then
$$ 
\int_0^\ell c(x,s)\,dx = \int_0^\ell c\left(\left[\left(Y_0(s) + \xi\right)\, \textrm{mod}\, \ell\right]\, \textrm{mod}\, 0,s\right) \,d\xi \quad \forall s \geq 0.
$$

The aim is to study a zero solution of Eq.~(\ref{eq:statphys4}) to obtain properties of the zero solution of Eq.~(\ref{eq:statphys8}). To do this we study the Laplace-Stieltjes transform $\mathscr{L}_S[|C(X(t),t)|]$ for $X(t) \in \bar{\kappa}_t$, 
where 
\begin{eqnarray*}
\bar{\kappa}_t = \{ X(t) \in \kappa_t \,|\, X(0)\in (I_A^+,I_B^-),\,\forall t >0 \quad \textrm{ either} \\
 C(X(t),t) > 0 , \quad C(X(t),t) < 0 \quad \textrm{or} \quad C(X(t),t) = 0  \}.
\end{eqnarray*}
The quantity $\mathscr{L}_S[|C(X(t),t)|]$ exists for every $X(t) \in \bar{\kappa}_t$; given $M(t)$ is smoothly defined through Eq.~(\ref{eq:statphys3}), we set $X_0(t) = M(t)$ in Eq.~(\ref{eq:origin}) and note that \\ $C\left(\left[\left(M(t) + \xi\right) \,\textrm{mod}\, \ell\right]\, \textrm{mod}\, 0,t\right)$ is differentiable a.e. in $[0, \infty)$ and therefore is of bounded variation. Provided
\begin{equation}\label{eq:statphysno}
\forall t \geq 0, \, M(t) \in (0, \ell) \quad \textrm{is the unique point in which} \quad C(M(t),t) = 0, 
\end{equation}
the kernel of $\mathscr{L}_S$, $\textrm{ker}[\mathscr{L}_S] = \left\{|C(X(t),t)| = k \,| \, X(t) \in \bar{\kappa}_t , k \geq 0 \right\} = C(M(t),t) = 0$. That is, if Eq.~(\ref{eq:statphysno}) is true, mass is exhausted out of the system until $\lim_{t \to \infty} C(x,t) = 0$ (since $\|J_i\| < \infty$), thereby ensuring that the only solution to $|C(X(t),t)| = k$ is $X(t) = M(t)$ when $k = 0$. If we take $\bar{\kappa}_s = \{ Y(s)\, | \, c(Y(s),s) = \mathscr{L}[C(X(t),t)], X(t) \in \bar{\kappa_t}\}$, then there exists a bijective map $\phi: \bar{\kappa}_t \to \bar{\kappa}_s$ such that if Eq.~(\ref{eq:statphysno}) is true, then for $X(t) \in \bar{\kappa}_t$,  $\mathscr{L}_S[|C(X(t),t)|]= |s\,c(\phi(X(t)),s) - C(X(0),0)|$ and  
\begin{equation}\label{eq:statphysno3}
 \forall s \geq 0,\,  \exists! y(s):  s\,c(y(s),s) = C(M(0),0),\quad \textrm{where} \quad \phi(M(t)) = y(s).
\end{equation}
As $M(0) \in (I^+_A, I^-_B)$, it follows that $c(y(s),s) = 0$. This result establishes the link between the two zero solutions $c(y(s),s)$ and $C(M(t),t)$. Note that such a link can be established through direct analysis of the $\mathscr{L}$ operator, although the above clearly shows how the mapping fails in the case in which the initial distributions are mixed (i.e. if $I^+_A = I^-_B$, then $C(M(0),0) \neq 0$ or is not defined).

It remains to show that $\partial_x C(M(t),t) = \mathscr{L}^{-1}[\partial_x c(y(s),s)]$. Re-define $X(t)\in \bar{\kappa}_t$ by $X(t) = M(t) + \xi$, so that $\forall \xi$, $(M(0) + \xi) \in (I^+_A,I^-_B)$. For some $|\xi|$ small, $\mathscr{L}:C(M(t) + \xi,t) \to c(\phi(M(t) + \xi),s)$ is a continuous map which ensures that $\phi$ is continuous and hence that
$$
\lim_{\xi \to 0} c(\phi(M(t) + \xi), s) = c(y(s),s) = 0 =  \mathscr{L}[C(M(t),t)] = \lim_{\xi \to 0} \mathscr{L}[C(M(t) + \xi,t)].
$$
From this, and Eq.~(\ref{eq:statphys3}) being well defined,  we have
$$
\frac{f(s)}{D} = \partial_x c(y(s),s) = \lim_{\xi \to 0}\partial_{\xi} c(\phi(M(t) + \xi),s) = 
\mathscr{L}[\lim_{\xi \to 0}\partial_{\xi} C(M(t) +\xi ,t)] =   \mathscr{L}\left[\frac{F(t)}{D}\right].
$$
It is therefore possible to define 
\begin{equation}\label{eq:statphys6}
f = f(s)^- = \lim_{x \to y(s)^-} - D\partial_x c \equiv  \lim_{x \to y(s)^+} \left|D \partial_x c \right|  = f(s)^+ = f. 
\end{equation}
This relation implies that $\partial_{xx} c(y(s),s) = 0$ and consequently, from Eq.~(\ref{eq:statphysno3}), that the left hands side of Eq.~(\ref{eq:statphys2}) is zero at $x =y(s)$. This is true provided $\forall s \geq 0, y(s) \notin [I^-_A, I^+_A]$ or $y(s) \notin [I^-_B, I^+_B]$. Conversely, if for any $s = s^*$ , $y(s) \in [I^-_A, I^+_A]$ or $y(s) \in [I^-_B, I^+_B]$ then Eq.~(\ref{eq:statphys3}) (and hence Eq.~(\ref{eq:statphysno})) would be false. By using this fact, the fact that  $c(I^+_A,s) >0$ and $c(I^-_B,s) < 0$ as $s \to \infty$, and a strong minimum principle~\cite[Pp 260]{Walter1998}, it can be shown that 
\begin{displaymath}
\forall s \geq 0, \quad  y(s) \in (I^+_A, I^-_B). 
\end{displaymath}
Through this condition, we are able to find a solution for $y(s)$ such that $\mathscr{L}[M(t)] \neq y(s)$; evaluating $x =y(s)$ in Eq.~(\ref{eq:statphys4}), setting  $c(y(s),s) = 0$, solving for $y(s)$ (by using the fact that $y(s) \in (I^+_A, I^-_B)$) results in Eq.~(\ref{eq:statphys9}). This can be used to find the required $f(s)$ through Eq.~(\ref{eq:statphys6}), the result being Eq.~(\ref{eq:statphys10}).

\begin{figure}
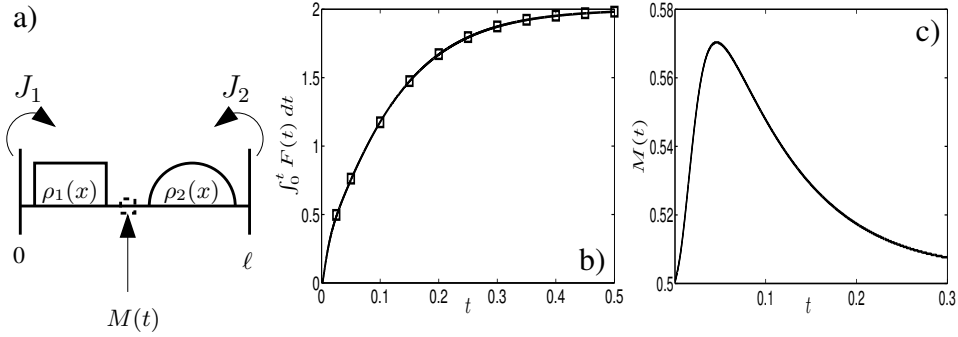
\caption{a) Illustration of problem. b) $\int_0^t F dt$ and c) $M(t)$ for the case $J_1 = 0$, $J_2 = - 20\,e^{-20 t}$, $\rho_A = \frac{20}{3}(u(x-\frac{1}{10}) - u(x - \frac{4}{10}))$, $\rho_B = k_1 \sin(\pi x)(u(x - \frac{6}{10}) - u(x - \frac{8}{10}))$, where $u(x)$ is the Heaviside step function. $M(t)$ is numerically determined using a Crank-Nicolson method~\cite{Press2007} and $\int_0^t F(t)\, dt$ is compared to the numerically inversed-Laplace transform of $s^{-1}f(s)$ (using~\cite{Stehfest1979}) as given by Eq.~(\ref{eq:statphys10}).}
\label{fig:statphys}
\end{figure}

\subsection{Uniqueness of $M(t)$}\label{sec:statphys2}
So far, it has been shown how to find the generating function of $F(t)$, the results being valid irrespective of the boundary conditions, provided $M(t)$ is unique. 
In this section, we define criteria (i) in $t$ or (i$^*$) and (ii$^*$) in $s$ for Eq.~(\ref{eq:statphysno}) to be true. 

For uniqueness of $M(t)$, we require $C(x,t) \lessgtr 0$ for $x \gtrless M(t)$ respectively, which implies that $C(0,t) \geq 0$, $C(\ell, t) \leq 0$ for all time. A boundary changing sign implies that there is either no moving Dirichlet homogeneous interior point condition or there is more than one. The case in which $C(0,t) = 0$ (say) on some time interval $T$ (with $C(x,t) < 0$ for $x > 0$) implies that the limiting $F(t)^-$ in Eq.~(\ref{eq:statphys3}) is not defined. We therefore require that  
\begin{enumerate}
\item{Neither $C(0,t)$ or $C(\ell,t)$ be identically zero for some $T$, unless $T = (0, \infty)$. For either case $j_1$ and $j_2$ cannot change signs.}
\end{enumerate}
Here, the case $T = (0, \infty)$ is considerable as homogeneous boundary conditions in $t$ are mapped to homogeneous boundary conditions in $s$. 

Denote $\mathcal{C}$ to be the set of completely monotonic functions~\cite{Berg2008}. A function $h(s) \in \mathcal{C}$, if 
\begin{equation}\label{eq:siamm}
h(s) = \mathscr{L}_S[H(t)], \quad (-1)^n\frac{d^n h}{ds^n} > 0, \quad n \geq 0, \quad s > 0,
\end{equation}
where $H$ is bounded and non-decreasing. 

It is possible to pose conditions in $s$, using the theory of completely monotonic functions. We require
\begin{enumerate}[(i*)]
\item $\forall\,s \geq 0$, $y(s) \in (I^+_A, I^-_B)$, 
\item $c(0,s)$, $-\,c(\ell,s)$ $\in \mathcal{C}$ unless $T = (0,\infty)$, in which case $j_1$, $-j_2$ $\in \mathcal{C}$.
\end{enumerate}
Further conditions can be derived from (i$^*$); taking $s = 0$ in $y(s)$ in Eq.~(\ref{eq:statphys9}) gives  $-j_1(0) - j_2(0) = 1 - v $ which implies that $\int_0^\ell C(x,0) \,dx = - \int_0^\infty J_1 + J_2 \,dt$. The latter is a mass balancing condition stating that the total mass leakage over time occurring at either side of the mixing point $M(t)$ must be equal. If this is not true, $M(t)$ converges to the boundary.

\subsubsection*{Example: Homogeneous Dirichlet boundary conditions.}
If $C(0,t) = C(\ell, t) = 0$, $v = 1$, $C(x,0) \neq C(\ell-x,0)$ then (i)  is not true.  To see this, consider
$C = C_3 = C_2 - C_1$, where $C_1(x,0) = \rho_A$, $C_2(x,0) = \rho_B$ and 
$$
\Delta C_i = \frac{\partial  C_i}{\partial t}  \quad \left\{ C_i(0,t) = C_i(\ell,t) = 0  \right\}, \quad i = 1,2. 
$$
If mass $Q_1$, $Q_2$, and $Q_3 = Q_2 - Q_1$ exits the system $1$,$2$ and $3$ respectively, then $\int_0^\infty Q_3\,dt =0$ and $Q_3$ must change sign unless $Q_1= Q_2$ which occurs when $C(x,0) = C(\ell-x,0)$.
Hence when, $C(x,0) \neq C(\ell-x,0)$, one of the boundary points is changing sign. Note for such examples (i$^*$) is satisfied.

\subsubsection*{Example: Homogeneous Neumann boundary conditions.}
If $v = 1$, $j_1 = j_2 = 0$, then (i) is true and the theory in \S~\ref{sec:statphys4} is applicable. If $v < 1$  then (i) is never true. In this case $M(t)$ takes some time $t_c$ to travels towards the $x = \ell$ boundary, such that $C(\ell, t_c) = 0$ and $C(x,t) > 0$ for all $x \in [0,\ell]$ for $t > t_c$. Here $t_c$ represents the time in which all species are mixed or the time at which there are no more $B$ species for the case in which $A$ and $B$ annihilate upon contact. 

\subsubsection*{Unphysical scenario.}
When a boundary flux changes signs, an unphysical situation occurs (i.e. $J_2(t)$ changing signs would imply a sudden introduction of $A$ species directly after an injection of $B$ species). For such a case the absolute mass in the system $\mathscr{M}(t)$ is not necessarily conserved. By considering the amount of mass in the system on either side of $M(t)$ it is found that
$$
\frac{d \mathscr{M}}{dt}\Big|_{x < M(t)} =  J_1(t) - F(t), \quad \frac{d \mathscr{M}}{dt}\Big|_{x > M(t)} =  J_2(t) -F(t),   
$$
where the subscripts denote the region of the domain.  If any mass introduced at the boundary exits the system at $x = M(t)$, and all mass drains out of the system as $t \to \infty$, then $ \|J_2\| + v - \| F\| = 0 = \|J_1\| + 1 - \| F\|$. If the latter condition is not true, then mass is not conserved throughout the system for all time. For example, changing $J_2  = -\frac{1}{20}\,e^{-20 t}\,+\,\sin(\frac{1}{10}\pi t)\,(1 - u(t-\frac{1}{5}))$ for the case in Fig.~\ref{fig:statphys}, results in $\| F(t)\| \neq \|J_2\| + v$ and implies that mass is being removed at some point $X(t) \neq M(t)$. Furthermore, it is possible to have cases in which $M(t)$ is defined for all time, with other mixing points forming within $(0,\ell)$ for a finite amount of time. For such situations, $f(s) = \mathscr{L}[F(t)]$, provided $M(t)$ is the unique point defined for all time.

\section{Multiple mixing points}~\label{sec:siam1}
In this section, we investigate the case where $A$ and $B$ are initially partitioned  over $N+1$ intervals in such a way that there are $N$ mixing points. The problem requires all odd or even intervals to be occupied by $A$ species ($A_1, A_3, \ldots$)  or $B$ species ($B_2, B_4, \ldots$) (respectively) such that a mixing point $M_i(t)$, $i = 1, \ldots, N$, forms from the interaction of the species in the $i$th interval $[I_i^-, I_i^+]$ and $(i+1)$th interval $[I_{i+1}^-, I_{i+1}^+]$ as illustrated in Fig~\ref{fig:statphys3}. The goal is to find under what conditions $M_i(t)$ will be defined for all time so as to determine whether the species are mixing homogeneously through $M_i(t)$.

\begin{figure}
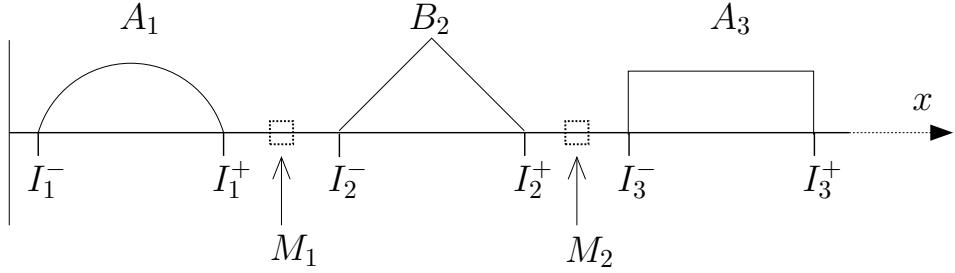
\caption{Illustration of multiple mixing points.}
\label{fig:statphys3}
\end{figure}

Through Lagrangian dynamics~\cite{Landau1976}, the variation of the energy in the {\it spatial variable} $x$ is found by studying the Lagrangian density  $\mathfrak{L} = \mathfrak{T}(x,s) - \mathfrak{P}(x,s)$, where $\mathfrak{T}$  is a kinetic energy and $\mathfrak{P}$ is a potential energy
such that
\begin{equation}
\mathfrak{T} = \left(D \partial_x c\right)^2 
\quad \textrm{and} \quad \mathfrak{P} = - s D\,c^2.
\end{equation}
If $M_i(t)$ is defined for all time then there must exist a $y_i(s)\in (I^+_i, I^-_{i+1}) \quad \forall s \geq 0$. At $x = y(s)$, $\mathfrak{P} = 0$ and $\sqrt{\mathfrak{T}} = f_i(s)$, which is the generating function of the mixing density at the $i$th mixing point. Because the mechanical energy produced in the interval is invariant in $x$,  the latter quantity can be found without determining $y_i(s)$. That is $\forall x \in (I^+_{i}, I^-_{i+1})$, $\mathfrak{T} + \mathfrak{P} = f_i^2 $, where
$$
f_i^2 = \left(Z_1\left[I^+_i, C(x,0) \right]+j_1\right)^2   - \frac{\left(\left(Z_1\left[I^+_i, C(x,0) \right]+j_1\right) \textrm{ch}(\ell) + Z_2\left[I^-_{i+1}, C(x,0)\right] + j_2\right)^2}{\textrm{sh}(\ell)^2}.
$$
To derive conditions concerning the existence of the mixing point for all time we use the fact that $f_i \in \mathcal{C}$ (or that $f_i(s) \notin \mathcal{C} \Rightarrow \not \exists M_i(t)\, \forall t >0$). As $f_i \in \mathcal{C} \Rightarrow f_i^2 \in \mathcal{C}$, we focus on $f_i^2$.

Taking a Taylor series expansions of $f_i^2$ about $s = 0$ results in
$$
f_i(s)^2 = \frac{a_1}{s} + a_2 + a_3 s + O(s^2), \quad \textrm{where} \quad a_i \in \mathbb{R}, i = 1, 2, 3. 
$$
As $f(0) = \int_0^\infty F(t)\,dt$ must exist, we require $a_1 = 0$. This amounts to setting $j_{{2}} \left( 0 \right) + j_1(0)  = -\int _{0}^{\ell}\!C \left( X,0 \right) {dX}$, which confirms the mass balancing statement described in \S~\ref{sec:statphys2}. By Eq.~(\ref{eq:siamm}), $a_3< 0$ and implies that 
\begin{equation}\label{eq:saimtor}
(\ell - R_i^+ - R_i^-)(\ell - R_i^+ + R_i^-)(\ell + R_i^+ - R_i^-)(\ell +R_i^+ + R_i^-) > 0
\end{equation}
where
\begin{eqnarray*}
R_i^+ = \sqrt{\frac{ \int_0^{I_i^+} x^2 C(x,0)\,dx - 2D\int_0^\infty t J_1\,dt}{ \int_0^{I_i^+} C(x,0)\,dx + \int_0^\infty J_1\,dt}} \quad \textrm{and} \\
R_i^- = \sqrt{\frac{ \int_{I_{i+1}^-}^\ell (\ell-x)^2 C(x,0) \,dx - 2D\int_0^\infty t J_2\,dt}{ \int_{I_{i+1}^-}^\ell C(x,0) \,dx + \int_0^\infty J_2 \,dt}},
\end{eqnarray*}
are signed radius of gyrations about the $x = 0$ and $x = \ell$ axis respectively. Given that $\ell > R_i^+$ and $\ell > R_i^-$, the only way the bound in Eq.~(\ref{eq:saimtor}) is satisfied is if
$$
\ell > R_i^+ + R_i^-.
$$
This provides a necessary condition for $M_i(t)$ to be defined for all time. Note that 
 when $J_1(t) = J_2(t) \equiv 0$, the above reduces to the result reported in Ref.~\cite{Haynes2012}. 

 
\section{Fractals}~\label{sec:siam2}
Understanding diffusion phenomenon through finitely ramified deterministic fractals continues to be of modeling importance to disordered media~\cite{Balankin2012, Balankin2013, Condamin2007}. To further our understanding of the mixing point on one dimensional domains, we consider the fluctuation density on a one dimensional path through-out a given fractal.

\begin{figure}
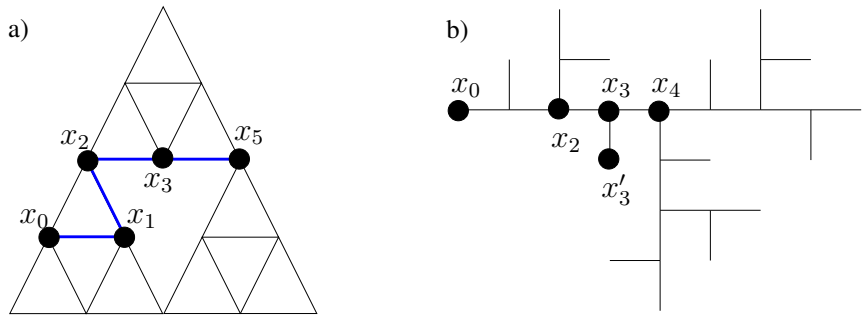
\caption{a) Illustration of a one dimensional path within a fractal. b) The third iteration of a T-tree fractal.}
\label{fig:statblorp}
\end{figure}

The analysis involves working with $C(x,t):  \Omega  \to \mathbb{R} $, $\Omega = [0, \ell_f] \times [0,\infty)$, is found to satisfy 
\begin{equation}\label{eq:statphys8b}
D \partial_{xx}\, C  = \partial_t C + D\sum_{i = 0}^N \delta (x - x_i) \partial_x C(x_i,t) \quad \textrm{and}  \quad C(x,0) = \rho_A(x) -\rho_B(x).
\end{equation}
Here $\ell_f$ is the length of the path throughout the fractal, $\rho_A$ and $\rho_B$ are the initial concentrations of $A$ and $B$ species {\it on the path} (other initial conditions can be specified throughout the fractal) and $x_i$ are points on the path where source fluctuation of the species can occur. For the example problem illustrated in Fig.~\ref{fig:statblorp} a), $\ell_f = 4\ell$, and fluctuations can occur at points $x_0$, $x_1$, $x_2$, $x_3$ or $x_4$. Note that boundary conditions at $x = 0$ and $x = x_f$ can be also be applied if required. Observe the difference between Eqs.~(\ref{eq:statphys8}) and (\ref{eq:statphys8b}); they both are one dimension problems, except Eq.~(\ref{eq:statphys8b}) has additional source terms at $x = x_i$. These source terms will affect the relation in Eq.~(\ref{eq:statphys3}). That is, an expression 
\begin{equation}\label{eq:statphys3f}
F(t) = F(t)^- = \lim_{x \to M(t)^-} - D\partial_x \,C \equiv  \lim_{x \to M(t)^+} \left| D\partial_x \,C \right|  = F(t)^+ = F(t),
\end{equation}
is not necessarily true given that fluctuations occurring at $x_i$  can make  such a result invalid. To further emphasis this point, we consider two examples to explain the behavior of the mixing point along a one dimensional path within a fractal.\\
\newline
Case i) There is a unique mixing point that remains between two source points on the path for all time; i.e. $\forall t > 0, M(t) \in [x_i, x_{i+1}] $. For example, when there is an instantaneous release of an equal number of $A$ and $B$ species at $x_2$ and $x_4$ (respectively) at time $t = 0$ on the T-tree illustrated in Fig.~\ref{fig:statblorp} b), with no-flux boundary conditions throughout the structure, $M(t)$ is always between $x_3$ and $x_4$. 

In general, the required $f(s)$ is derived as in \S~\ref{sec:siam1} by studying the energy $\mathfrak{T}+ \mathfrak{P} = f^2$ produced in the interval $(x_{i}, x_{i+1})$.  It is found that 
\begin{equation}\label{eq:siammn}
\sqrt{\mathfrak{P} + \mathfrak{T}} = \sqrt{s}\frac{\sqrt{p_{x_i}^2 + p_{x_{i+1}}^2 - 2 p_{x_i} p_{x_{i+1}}\textrm{ch}(\ell)}}{\textrm{sh}(\ell)} = f,
\end{equation}
where $p_{x_i}$ and $p_{x_{i+1}}$ are the Laplace transformed concentrations at $x_i$ and $x_{i+1}$ (and can be found through Ref.~\cite{Haynes2008}). This provides a means of testing results on fractals and developing asymptotic scaling laws.  \\
\newline
Case ii) There is a unique mixing point that does not remain between two source points on the path for all time. In this case, it is not possible to calculate the mixing density. To see this, consider the instantaneous release of an equal number of $A$ and $B$ species at $x_0$ and $x_4$ at time $t = 0$ on the T-tree illustrated in Fig.~\ref{fig:statblorp} b), with no-flux boundary conditions throughout the structure. For the system in $t$, a unique mixing point $M(t)$ will originate at $x_2$ at $ t= 0$ and will then move towards $x_4$. When it arrives at $x_3$,  $M(t)$ will branch into two mixing points, one that moves up into the domain $[x_3, x_3']$ and the other which continue towards $x_4$. This violates the uniqueness of $M(t)$. For the system in $s$, there is a unique moving homogeneous Dirichlet point, $y(s)$, that originates at $x_2$ as $s \to \infty$ and converges between $x_3$ and $x_4$ as $s \to 0$. When $y(s)$ is between points $x_2$ and $x_3$ or $x_3$ and $x_4$, its value is correctly defined and is such that there exists some $s_{*}$ such that $\lim_{s \to s_{*}^-} y(s) =  \lim_{s \to s_{*}^+} y(s) = x_3$. However, there are two different formulations for $y(s)$ depending on the interval $y(s)$ lies in and the theory breaks down.  Note it is possible that through a particular choice of initial conditions the above problem involves a unique formulation for $y(s)$ over both intervals for all $s \geq 0$. If this occurs, then the analysis is similar to case i) and exact results can be found for $f(s)$.

Even if there are no exact results for this problem, it is hypothesized that if a unique homogeneous Dirichlet point $y(s)$ converges to $y_f(s)$ defined in some domain $[x_f, x_{f+1}]$ as $s \to 0$ then the resulting $f_f(s)$ should give an approximation to the long time mixing behavior of $A$ and $B$ i.e. $F(t) \approx \mathscr{L}^{-1}[f_f(s)]$, where $F(t)$ is the true mixing density. Indeed,  the difference between the two examples is the additional mixing point that originate in the side branch in case ii). This mixing point dies off rather quickly and the asymptotic mixing behavior of  $A$ and $B$ within the structure for large $t$ should be dominated by the same mechanism as Eq.~(\ref{eq:siammn}). 

\section{Summary and Discussion}\label{sec:statphysend}
The mixing time generating function for the diffusion reaction system $A+B\to 0$ on a one dimensional domain was derived for the case in which homogeneous mixing occurred through a unique mixing point for general initial and boundary conditions. The mixing time represents the reaction time for the special case in which $A$ and $B$ annihilate upon contact. Note that the study was restricted to continuous initial and boundary conditions as it realistically coincides with experimental conditions, although it is possible to relax these conditions. 

The generalization and application of this method to modified diffusion-reaction schemes involving more general evolution equations would be of interest. This does not make reference to formulations $A + B \to B$ where $B$ is a moving trap, or the reaction $A + B \to C$ whose results can analogously be interpreted from the result presented within, but to modified problems  such as the reaction- sub-diffusion problem~\cite{Henry2002}, or cases which require further interior point conditions (e.g. semi-permeable cellulose membrane~\cite{PhysRevE.75.026107}).

From this study, a natural question to ask is whether $F$ can be used to define the reaction rate through a form $R(\xi,t) = \int_0^t F(\tau)\,G(\xi,t-\tau)\,d\tau$, for some specified $G(\xi,t)$ whose physical significance remains unknown. Indeed for the case in which $A$ and $B$ annihilate upon contact at $x = M(t)$ ($\xi = 0$), $G(0,t) = \delta(t)$ and $R(M(t),t) = F(t)$. Further motivation for defining $R(x,t)$ in terms of $F(t)$ lies in the case in which $D_A \neq D_B$; as the analysis presented in the beginning of section \S~\ref{sec:statphys} to Eq.~(\ref{eq:statphysno3}) can be applied to the case $C(x,t) = D_A C_A - D_B C_B$, the analysis of the homogeneous Dirichlet boundary conditions is still possible, although $F(t)$ will ultimately depend on $R(M(t),t)$. This places emphasis on the fact that the particles mixing behavior will ultimately depend on how they react when they first encounter one another, a modeling aspect that  marries well with the introductory statement to this paper. 

The theory presented was extended to consider multiple mixing points and general one dimensional systems (i.e. finitely ramified fractals). We derived a necessary condition to determine whether $A$ and $B$ were mixing homogeneously through a given mixing point. It was shown that for finitely ramified fractals, the mixing density can be governed by two different mechanisms; depending on the initial placement of the $A$ and $B$ species, a purely one dimensional mixing mechanism can occur. This becomes significant when dealing with fractals with no loops. It is believed that further progress in deriving the mixing density in fractal domains can be made using the respective propagator for fractals~\cite{PhysRevLett.54.455}.

\end{document}

%% file: paperfig.pdf_t
\setlength{\unitlength}{2368sp}%

\begingroup\makeatletter\ifx\SetFigFont\undefined%
\gdef\SetFigFont#1#2#3#4#5{%
  \reset@font\fontsize{#1}{#2pt}%
  \fontfamily{#3}\fontseries{#4}\fontshape{#5}%
  \selectfont}%
\fi\endgroup%
\begin{picture}(10201,3612)(4350,-4573)
\put(4726,-3136){\makebox(0,0)[lb]{\smash{{\SetFigFont{9}{10.8}{\rmdefault}{\mddefault}{\updefault}{\color[rgb]{0,0,0}$\rho_1(x)$}%
}}}}
\put(5401,-4486){\makebox(0,0)[lb]{\smash{{\SetFigFont{9}{10.8}{\rmdefault}{\mddefault}{\updefault}{\color[rgb]{0,0,0}$M(t)$}%
}}}}
\put(6001,-3136){\makebox(0,0)[lb]{\smash{{\SetFigFont{9}{10.8}{\rmdefault}{\mddefault}{\updefault}{\color[rgb]{0,0,0}$\rho_2(x)$}%
}}}}
\put(6826,-3886){\makebox(0,0)[lb]{\smash{{\SetFigFont{9}{10.8}{\rmdefault}{\mddefault}{\updefault}{\color[rgb]{0,0,0}$\ell$}%
}}}}
\put(4426,-2086){\makebox(0,0)[lb]{\smash{{\SetFigFont{12}{14.4}{\rmdefault}{\mddefault}{\updefault}{\color[rgb]{0,0,0}$J_1$}%
}}}}
\put(6601,-2086){\makebox(0,0)[lb]{\smash{{\SetFigFont{12}{14.4}{\rmdefault}{\mddefault}{\updefault}{\color[rgb]{0,0,0}$J_2$}%
}}}}
\end{picture}%

\vspace{- 5 mm}
\begin{picture}(0,0)%
\includegraphics{paperfig.pdf}%
\end{picture}%

%% file: example2.pdf_t
\begin{picture}(0,0)%
\includegraphics{example2.pdf}%
\end{picture}%
\setlength{\unitlength}{789sp}%
\begingroup\makeatletter\ifx\SetFigFont\undefined%
\gdef\SetFigFont#1#2#3#4#5{%
  \reset@font\fontsize{#1}{#2pt}%
  \fontfamily{#3}\fontseries{#4}\fontshape{#5}%
  \selectfont}%
\fi\endgroup%
\begin{picture}(29727,8489)(889,-5918)
\put(7300,-3365){\makebox(0,0)[lb]{\smash{{\SetFigFont{14}{16.8}{\rmdefault}{\mddefault}{\updefault}{\color[rgb]{0,0,0}$I^+_1$}%
}}}}
\put(1479,-3365){\makebox(0,0)[lb]{\smash{{\SetFigFont{14}{16.8}{\rmdefault}{\mddefault}{\updefault}{\color[rgb]{0,0,0}$I^-_1$}%
}}}}
\put(4389,1728){\makebox(0,0)[lb]{\smash{{\SetFigFont{14}{16.8}{\rmdefault}{\mddefault}{\updefault}{\color[rgb]{0,0,0}$A_1$}%
}}}}
\put(20032,-3365){\makebox(0,0)[lb]{\smash{{\SetFigFont{14}{16.8}{\rmdefault}{\mddefault}{\updefault}{\color[rgb]{0,0,0}$I^-_3$}%
}}}}
\put(25855,-3365){\makebox(0,0)[lb]{\smash{{\SetFigFont{14}{16.8}{\rmdefault}{\mddefault}{\updefault}{\color[rgb]{0,0,0}$I^+_3$}%
}}}}
\put(22943,1728){\makebox(0,0)[lb]{\smash{{\SetFigFont{14}{16.8}{\rmdefault}{\mddefault}{\updefault}{\color[rgb]{0,0,0}$A_3$}%
}}}}
\put(10938,-3365){\makebox(0,0)[lb]{\smash{{\SetFigFont{14}{16.8}{\rmdefault}{\mddefault}{\updefault}{\color[rgb]{0,0,0}$I^-_2$}%
}}}}
\put(16759,-3365){\makebox(0,0)[lb]{\smash{{\SetFigFont{14}{16.8}{\rmdefault}{\mddefault}{\updefault}{\color[rgb]{0,0,0}$I^+_2$}%
}}}}
\put(9119,-5546){\makebox(0,0)[lb]{\smash{{\SetFigFont{14}{16.8}{\rmdefault}{\mddefault}{\updefault}{\color[rgb]{0,0,0}$M_1$}%
}}}}
\put(18397,-5593){\makebox(0,0)[lb]{\smash{{\SetFigFont{14}{16.8}{\rmdefault}{\mddefault}{\updefault}{\color[rgb]{0,0,0}$M_2$}%
}}}}
\put(29311,-818){\makebox(0,0)[lb]{\smash{{\SetFigFont{14}{16.8}{\rmdefault}{\mddefault}{\updefault}{\color[rgb]{0,0,0}$x$}%
}}}}
\put(13484,1728){\makebox(0,0)[lb]{\smash{{\SetFigFont{14}{16.8}{\rmdefault}{\mddefault}{\updefault}{\color[rgb]{0,0,0}$B_2$}%
}}}}
\end{picture}%

%% file: sierp2.pdf_t
\begin{picture}(0,0)%
\includegraphics{sierp2.pdf}%
\end{picture}%
\setlength{\unitlength}{789sp}%
\begingroup\makeatletter\ifx\SetFigFont\undefined%
\gdef\SetFigFont#1#2#3#4#5{%
  \reset@font\fontsize{#1}{#2pt}%
  \fontfamily{#3}\fontseries{#4}\fontshape{#5}%
  \selectfont}%
\fi\endgroup%
\begin{picture}(26831,9709)(1139,-5194)
\put(15060,1800){\makebox(0,0)[lb]{\smash{{\SetFigFont{12}{14.4}{\rmdefault}{\mddefault}{\updefault}{\color[rgb]{0,0,0}$x_0$}%
}}}}
\put(18210, 75){\makebox(0,0)[lb]{\smash{{\SetFigFont{12}{14.4}{\rmdefault}{\mddefault}{\updefault}{\color[rgb]{0,0,0}$x_2$}%
}}}}
\put(21360,1800){\makebox(0,0)[lb]{\smash{{\SetFigFont{12}{14.4}{\rmdefault}{\mddefault}{\updefault}{\color[rgb]{0,0,0}$x_4$}%
}}}}
\put(19785,1800){\makebox(0,0)[lb]{\smash{{\SetFigFont{12}{14.4}{\rmdefault}{\mddefault}{\updefault}{\color[rgb]{0,0,0}$x_3$}%
}}}}
\put(19785,-1500){\makebox(0,0)[lb]{\smash{{\SetFigFont{12}{14.4}{\rmdefault}{\mddefault}{\updefault}{\color[rgb]{0,0,0}$x_3'$}%
}}}}
\put(1501,-2311){\makebox(0,0)[lb]{\smash{{\SetFigFont{12}{14.4}{\rmdefault}{\mddefault}{\updefault}{\color[rgb]{0,0,0}$x_0$}%
}}}}
\put(4876,-2311){\makebox(0,0)[lb]{\smash{{\SetFigFont{12}{14.4}{\rmdefault}{\mddefault}{\updefault}{\color[rgb]{0,0,0}$x_1$}%
}}}}
\put(5401,-1186){\makebox(0,0)[lb]{\smash{{\SetFigFont{12}{14.4}{\rmdefault}{\mddefault}{\updefault}{\color[rgb]{0,0,0}$x_3$}%
}}}}
\put(2776,164){\makebox(0,0)[lb]{\smash{{\SetFigFont{12}{14.4}{\rmdefault}{\mddefault}{\updefault}{\color[rgb]{0,0,0}$x_2$}%
}}}}
\put(8251,314){\makebox(0,0)[lb]{\smash{{\SetFigFont{12}{14.4}{\rmdefault}{\mddefault}{\updefault}{\color[rgb]{0,0,0}$x_5$}%
}}}}
\end{picture}%